\documentclass[aps,pre,showkeys,twocolumn,longbibliography]{revtex4-2}
\usepackage{graphicx}
\usepackage{dcolumn}
\usepackage{bm}
\usepackage{amsmath}
\usepackage{amssymb}
\usepackage{float}
\usepackage{xcolor}

\begin{document}
\title{Two-field theory for phase coexistence of active Brownian particles}

\author{Pablo Pérez-Bastías}
\author{Rodrigo Soto}
\affiliation{Departamento de Física, FCFM, Universidad de Chile, Blanco Encalada 2008, Santiago, Chile}

\begin{abstract}
Active Brownian particles (ABPs) serve as a minimal model of active matter systems. When ABPs are sufficiently persistent, they undergo a liquid-gas phase separation and, in the presence of obstacles, accumulate around them, forming a wetting layer. Here, we perform simulations of ABPs in a quasi-one-dimensional domain in the presence of a wall, studying the dynamics of the polarization field. On the course of time, we observe a transition from a homogeneous (where all particles are aligned) to a heterogeneous (where particles align only at the interface) polarization regime. We propose coarse-grained equations for the density and polarization fields based on microscopic and phenomenological arguments that correctly account for the observed phenomena.
\end{abstract}

\maketitle

Active matter, composed of interacting individuals capable of self-propel~\cite {ramaswamy_active_2017},  is usually studied using agent models such as run-and-tumble particles, Ornstein--Uhlenbeck particles~\cite{martin_statistical_2021}, and active Brownian particles (ABPs)~\cite{bechinger_active_2016}. The last model, the focus of this letter, has become a prototypical theoretical model of active matter because of its simplicity. In addition, it is relevant in describing systems composed of smooth swimmers such as some types of bacteria, microalgae, or active colloids~\cite{bechinger_active_2016} and, for collective phenomena, ABPs are equivalent to run-and-tumble particles~\cite{cates_when_2013}. ABPs move with constant speed $v_0$, whose direction angle $\theta$ changes by angular diffusion with a rotational diffusion coefficient $D_r$. The dynamics of the particle $i$ in two spatial dimensions is governed by the equations
\begin{align}\label{eq:vor_abp}
    \dot{\mathbf{r}}_i &= v_{0} \mathbf{n}_i + \mathbf{F}_{i}, &
    \dot{\theta}_i &= \sqrt{2D_r} \eta_i(t),
\end{align}
where the director $\mathbf{n}_i=\left(\cos{\theta_i}, \sin{\theta_i}\right)$ determines the direction of the self-propulsion and $\eta_i$ is a Gaussian white noise with zero mean and $\langle \eta_i(t) \eta_j(t') \rangle = \delta_{ij}\delta(t - t')$. In addition, $\mathbf{F}_i = -\sum_{i \neq j} \nabla U\left(|\mathbf{r}_i - \mathbf{r}_j|\right)$ is the net force on particle $i$ due to interactions with other particles $j$, mediated by a potential $U(r)$. Usually, $U$ is an excluded-volume potential, with particles of characteristic size $\sigma$. As usual in overdamped systems, the friction coefficient is absorbed into the force, which therefore has units of velocity. Note that the particles move a distance $\ell = v_0 / D_r$ before changing direction; this quantity is called the persistent length. Here, we will consider the regime of high persistence, in which case any translational noise is subdominant in comparison to the active diffusivity that scales as $v_0^2/D_r$~\cite{berg2025random}. Hence, for simplicity, no translational noise is considered in the model. 

Even if the interparticle interactions are purely repulsive, when ABPs are highly persistent ($D_r \ll v_0 / \sigma$), they undergo a liquid-gas  separation, called the motility-induced phase separation (MIPS)~\cite{cates_motility-induced_2015}. 
Similarly, in the presence of obstacles, ABPs accumulate around them, forming a wetting layer~\cite{sepulveda_wetting_2017, sepulveda_universality_2018,turci2021wetting,neta_wetting_2021}, as it has been observed experimentally for active colloids~\cite{fins_carreira_how_2024} and with inertial active particles~\cite{caprini2024dynamical}. The origin of both phenomena is that ABPs, as a result of interactions, move with an effective velocity $v(\rho)$, which is a decreasing function of the density $\rho$~\cite{cates_diffusive_2012,soto_self-diffusive_2025}. This velocity reduction provides the mechanism for wall accumulation and MIPS~\cite{cates_motility-induced_2015}. 

Continuum field theories for the density field can be derived using the Smoluchowski equation associated with Eqs.~\eqref{eq:vor_abp}~\cite{bialke_microscopic_2013}. To this end, the effective velocity is modeled as $v(\rho) = v_0 (1 - \rho / \rho_l)$ \cite{cates_diffusive_2012}, which vanishes upon approaching the liquid density $\rho_l$. The polarization field is then approximated as being proportional to the local density gradient, $\mathbf{q} \approx - \tfrac{1}{2 D_r} \nabla [v(\rho) \rho]$ \cite{bialke_microscopic_2013}. Under this adiabatic slaving approximation, an effective Cahn–Hilliard equation is obtained for the density field~\cite{speck_effective_2014, speck_dynamical_2015, speck_critical_2022}. Since active systems violate detailed balance, non-variational terms can also be included in the dynamical equations~\cite{stenhammar_continuum_2013, wittkowski_scalar_2014}. 

It should be noted that in the limit of high persistence, $D_r \to 0$, the local-density-gradient approximation fails to capture the system dynamics. First, in this regime the polarization field relaxes on long timescales. Hence, for intermediate time scales, it cannot be slaved to the density field and should be considered as a slow field together with density. Second, this approximation can lead to diverging solutions for $\mathbf q$ at interfaces, where density gradients and the effective velocity are non-zero~\cite{caprini_spontaneous_2020,hermann2020active,rojas-vega_wetting_2023}. 
Finally, for thin wetting layers, particles will be almost completely polarized against the wall, 
independently of the density profiles, and the gradient approximation breaks down. This is a bias effect because particles that reach the film are necessarily polarized and they remain so for a long time, and when they eventually point out of the wall they escape~\cite{lee_active_2013, elgeti_wall_2013}. This is related to the anisotropy in the motion of ABPs for intermediate time scales~\cite{basu_active_2018}.
These results, indicate that for high persistence, the system dynamics should, at least,  be described by the density and polarization fields together.

In this article, we derive continuum coarse-grained equations for the density and polarization fields of highly persistent interacting ABPs. For this purpose, we use the Dean method~\cite{dean_langevin_1996}, which is fed with phenomenological expressions taken from agent-based simulations. The resulting equations correctly describe the simulations results at different timescales. 

\begin{figure*}
    \centering
    \includegraphics[width=2.0\columnwidth]{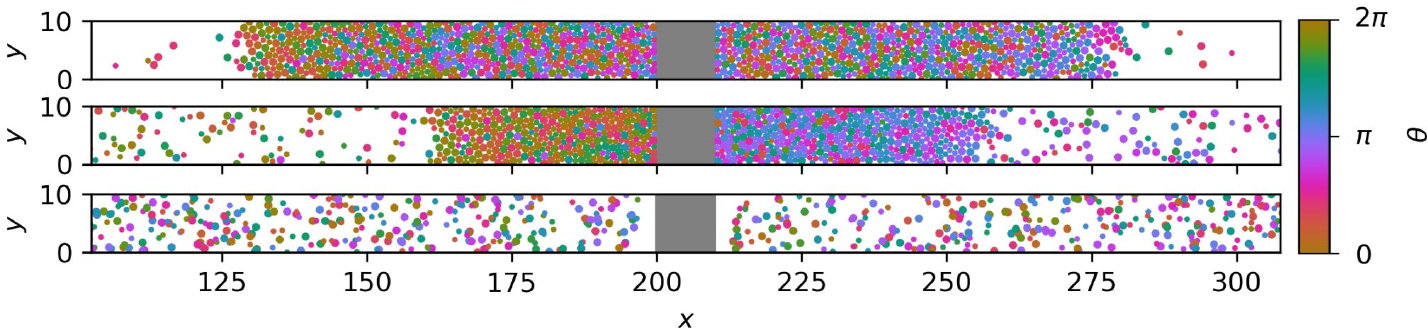}
    \caption{Snapshots of one of the simulations for $t = 0$ (bottom), 500 (middle), and 5000 (top). The color indicates the particle orientation angle $\theta$. Only the central part in $x$ of the system is shown.}
    \label{fig:snapshot}
\end{figure*}

\section{Simulations}\label{sec.simulations}

The Dean method or, in fact, any method that starts from the microscopic equations of motion to derive coarse-grained equations, necessarily needs to introduce approximate closures at some moment~\cite{soto2016kinetic,livi2017nonequilibrium,sarracino2025nonequilibrium}. Here, we use numerical simulations of the ABP model to obtain a phenomenological closure of the equations. With this purpose in mind, we will consider a quasi one-dimensional geometry in presence of rigid walls where, starting from an homogeneous state, the particles accumulate in the walls forming wetting layers (see Fig.~\ref{fig:snapshot}). The advantage of this geometry is that the relevant fields depend only on one coordinate and present both transient and stationary regimes, which will allow us to obtain the non-equilibrium fluxes needed to close the equations.

We perform sixteen simulations of $N=1200$ ABPs moving in two dimensions, interacting via a Weeks-Chandler-Andersen (WCA) like potential (see Ref.~\cite{rojas-vega_mixtures_2023} for details) to avoid overlapping. Half of the particles have diameter $\sigma =1$, and the other half have diameter $1.4 \sigma$, to prevent crystalline order to form in the dense phase. The self-propulsion speed is $v_0 = 1$ and, the rotational diffusion coefficient $D_r = 10^{-3} v_0 / \sigma$. The dimensions of the domain are $L_x=410\sigma$ and $L_y=10 \sigma$, with periodic boundary conditions in both directions. The particles move in the presence of a fixed wall along the short axis of the box, of width $10 \sigma$, located at $x=200 \sigma$. Due to the high persistence, particles can eventually travel ballistically from one wall to another. Finally, the simulation time is $T=5\times 10^{3} v_0 / \sigma$, with a simulation time step $\Delta t = 5\times 10^{-4} v_0 / \sigma$. Snapshots of one simulation are shown in Fig.~\ref{fig:snapshot} at different time instants. Initially, particles are distributed with random positions and orientations, respecting the excluded volume condition. In the course of time, they accumulate on both sides of the wall, reaching a steady state with permanent wetting films.

\begin{figure*}
 \includegraphics[width=2.0\columnwidth]{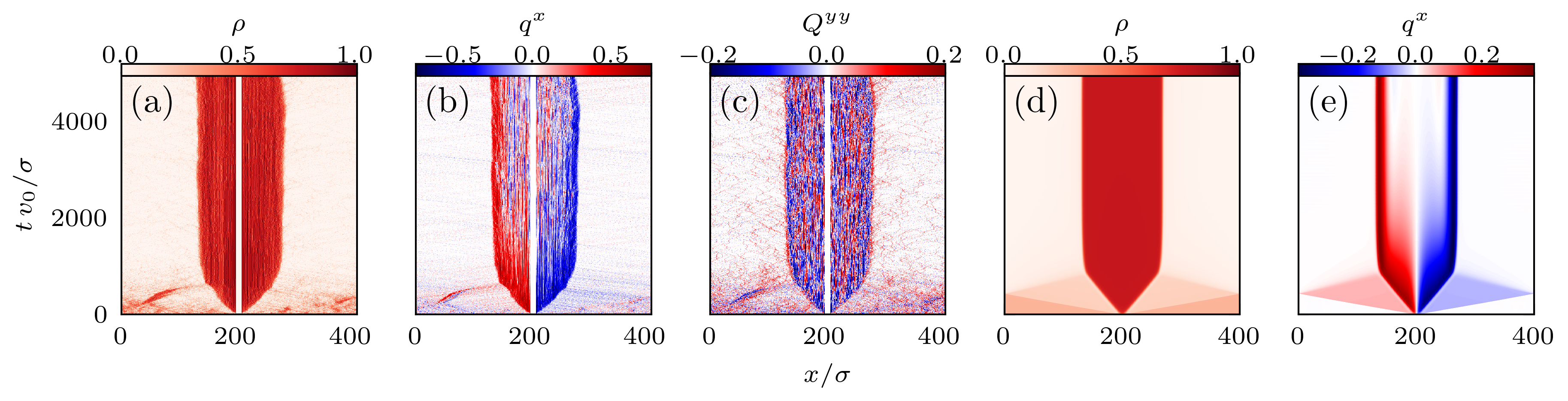}
 \includegraphics[width=2.0\columnwidth]{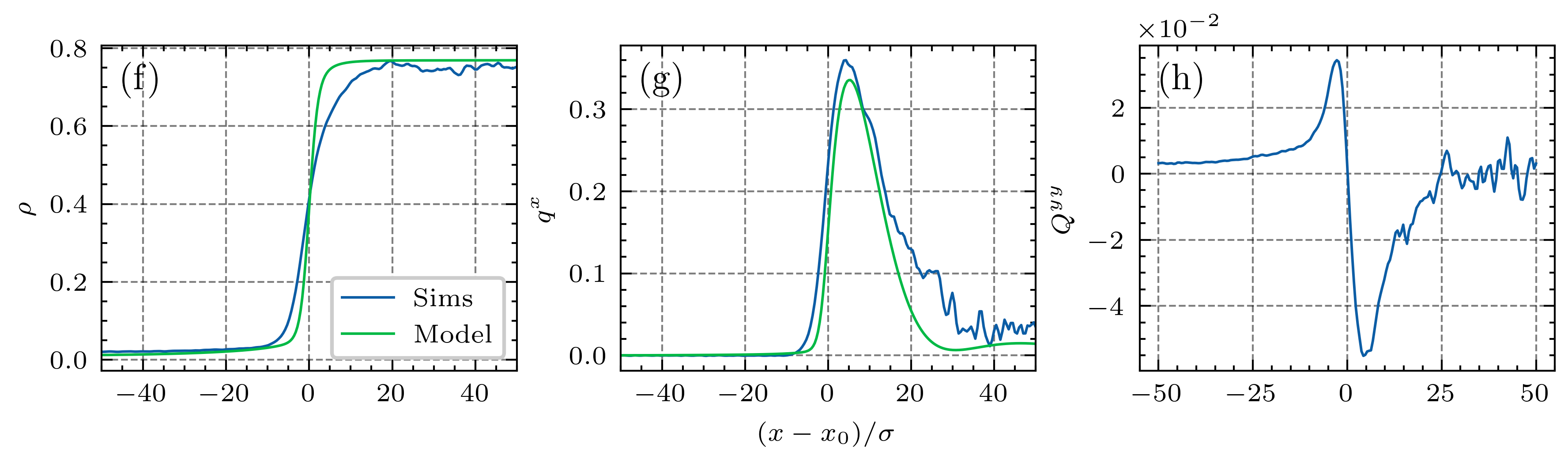}
    \caption{Top: Spatiotemporal diagrams 
    obtained from simulations for the (a) density, (b) $x$ component of polarization, $q^x$, and (c) $yy$ component of the nematic tensor, $Q^{yy}$. Spatiotemporal diagrams for density (d) and polarization (e) fields obtained from the numerical solutions of Eqs.~\eqref{eq:rho_approx} and \eqref{eq:qx_approx}. Bottom: Steady state profiles of the (f) density, (g) $x$ component of polarization, $q^x$, and (h) $yy$ component of the nematic tensor, $Q^{yy}$ of one interface, centered at $x_0$, with the dilute phase at the left and the dense at the right (see text for details). The blue lines are the results obtained from simulations and the green lines are obtained by solving numerically Eqs.~\eqref{eq:rho_approx} and~\eqref{eq:qx_approx}.}
    \label{fig:fields}
\end{figure*}

We compute the density $\rho$, the polarization $q^{x} = \rho\langle \cos \theta \rangle$, and the $yy$ component of the nematic tensor $Q^{yy} = -\tfrac{\rho}{2}\langle \cos(2\theta) \rangle$ through numerical coarse-graining. The spatiotemporal $x$–$t$ diagrams of these fields, for a representative simulation, are shown in Figs.~\ref{fig:fields}(a-c). Similar spatiotemporal diagrams were obtained for the other simulations. For each time step, we average along the $y$-axis. The stationary profile of all fields is computed by averaging over $t > 2500 \sigma / v_0$. 
When the wetting layers are formed, the total number of particles in the condensed phase is constant except for fluctuation. But, the two layers can change freely their width, and only their sum is fixed. This evolution is slow, resulting in stationary profiles for each simulation but with different interface positions among simulations. To average these profiles, we locate the center position of each interface $x_0$ by fitting the density profile to an hyperbolic tangent. 
Then, all profiles are aligned with $x_0$, reflecting in $x$ the profiles at the dense-to-dilute interfaces.
This allow us to obtain  unique stationary profiles, shown in Fig.~\ref{fig:fields}-bottom. 

As shown in Figs.~\ref{fig:fields}(b), (c), (g), and (h), despite the absence of aligning interactions between particles, the polarization and the nematic tensor are non-zero in interfacial regions. In the outer region of the interface, $Q^{yy}$ is positive. This indicates that, on average, the particles in this region move tangentially to the interface. This result has been reported before and is proposed as a mechanism for interface stabilization~\cite{omar_microscopic_2020, lee_interface_2017}. In the zones of the peaks of the polarization and the negative peaks of $Q^{yy}$, the particles point toward the wall.

\begin{figure}
 \includegraphics[width=\columnwidth]{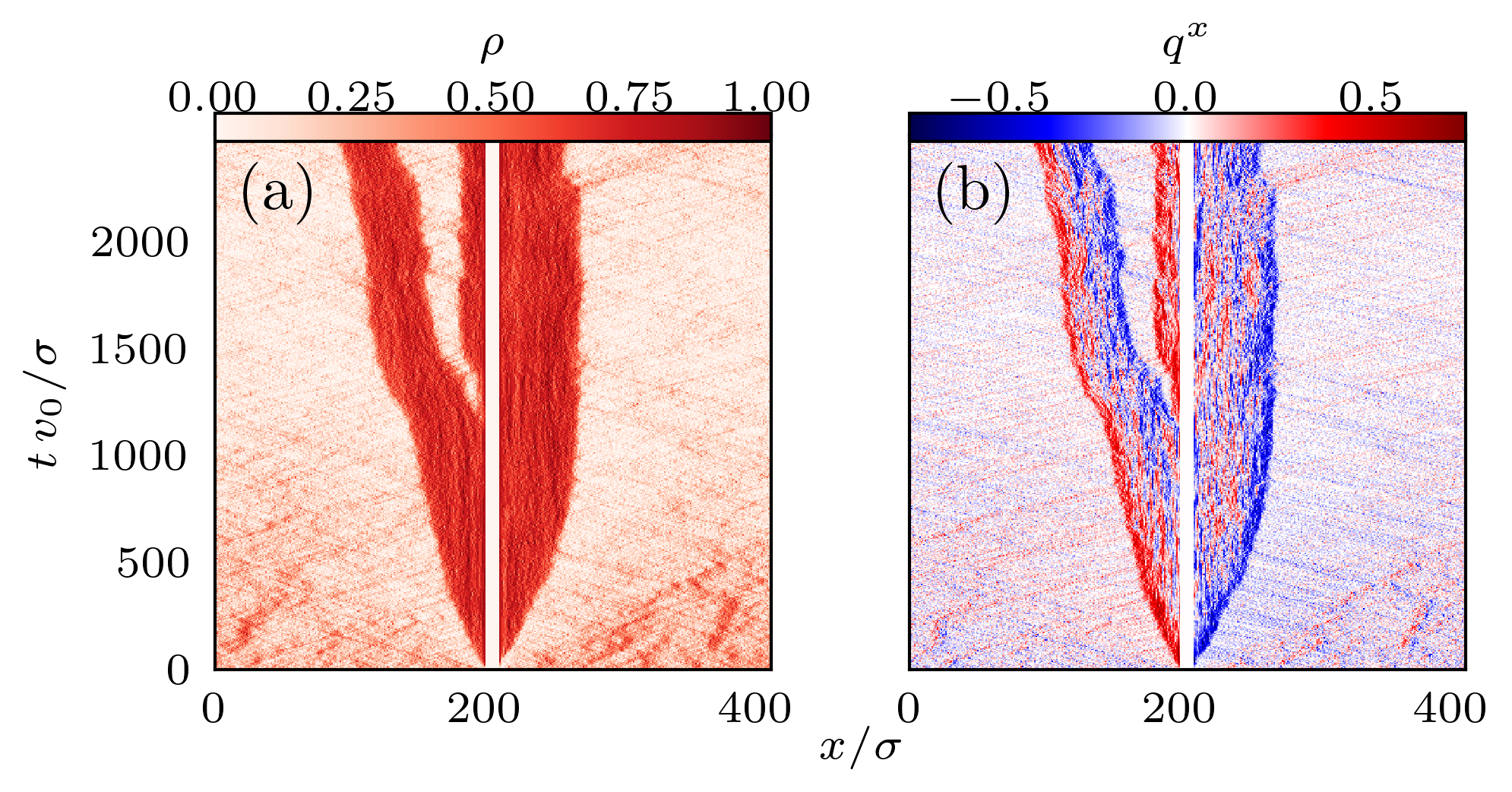}
 \caption{Spatiotemporal diagrams 
    obtained from a simulation with $D_1=0.01$, with the rest of the parameters equal to the other simulations. (a) shows the density and (b) the $x$ component of polarization, $q^x$.}
    \label{fig:fields2}
\end{figure}

Due to the high persistence, particles can travel ballistically from one wall to another. This because $L_x < \ell$. The value of the ratio $ K = L_x / \ell$ classifies different polarization signatures. When $K < 1$, as in the present case for which $K = 0.4$, there is a transition between two regimes. First, for $t < 1 / D_r$, all particles in the wetting film align towards the wall. This results in a homogeneous polarization state, shown in Fig.~\ref{fig:fields}(b) (also seen in Fig.~\ref{fig:snapshot}-middle). This is a result of the anisotropic dynamics at short timescales given by the persistent movement in the direction to which they were initially pointing~\cite{basu_active_2018}. This almost ballistic motion is appreciated as diagonal lines crossing the box length in the spatio-temporal diagrams~\cite{rojas-vega_mixtures_2023}. Second, for $t > 1/D_r$, there is a non-zero polarization only at the interface in a heterogeneous polarization regime (also seen in Fig.~\ref{fig:snapshot}-top). This permanent polarization originates microscopically by a bias effect, as particles pointing toward the dense phase remain blocked, while those pointing away can escape. Hence, for the homogeneous (heterogeneous) regime, the polarization field does not (does) follow the density gradient approximation. For $K> 1$, as particles change direction before reaching the wall, only the heterogeneous regime is observed. This is shown in Fig.~\ref{fig:fields2} (top), where the spatio-temporal diagrams of a simulation with $D_r = 10^{-2} v_0 / \sigma$ is presented. For this case, $K = 4$. The rest of the parameters are the same as the simulation with $D_r = 10^{-3} v_0 / \sigma$. Note that for this case, a part of the wetting layer  detaches due to fluctuations. To reproduce the detachment process, it would be necessary to consider the effects of the multiplicative noise in the field equations.

\section{Dynamical equations}
\subsection{Exact equations}
Taking note of the results of the simulations, we proceed to derive coarse-grained equations for the density and polarization fields valid for the two regimes described above. The dynamic of the nematic tensor is not considered, as its magnitude is smaller than the other two fields [see Fig.~\ref{fig:fields}(c)]. We follow the Dean method~\cite{dean_langevin_1996, nakamura_derivation_2009} since it provides insights on the system noise, which could be relevant for the overall dynamics~\cite{keta_collective_2021, han_fluctuating_2021, feliachi_fluctuating_2022}, as we seen in Fig.~\ref{fig:fields2}. 
We define the global density, polarization, and nematic fields
\begin{align}
    \rho(\mathbf{r}, t) &= \sum_i \delta(\mathbf{r}_i - \mathbf{r}),\\
    q^{\alpha}(\mathbf{r}, t) &= \sum_{i}n^{\alpha}_i  \delta(\mathbf{r}_i - \mathbf{r}),\\
    {Q}^{\alpha \beta}(\mathbf{r}, t) &=\sum_{i} \delta(\mathbf{r}_i - \mathbf{r}), t)\left[ {n}^{\alpha}_i {n}^{\beta}_i  - \frac{\delta^{\alpha \beta}}{2}\right],
\end{align}
where $\alpha$ and $\beta$ indicate the Cartesian components of the vectors and tensors, and the sums run over the particles $i$. Using the  microscopic equations \eqref{eq:vor_abp}, the Dean method gives (see Appendix~\ref{app.dean})
\begin{align}
    {\partial_t \rho} &= - v_0 \nabla_{\alpha} \bigg[ q^{\alpha} + G^{\alpha} \bigg], \label{eq:rho_micro} \\
    {\partial_t q^{\alpha}} &= - D_r q^{\alpha} -  v_0\nabla_{\beta} T^{\alpha \beta} + \sqrt{2 D_r \rho} \mu^{\alpha}, \label{eq:q_micro}
\end{align}
where $G^{\alpha} = \frac{1}{v_0}\sum_i F^{\alpha}_i \delta(\mathbf{r}-\mathbf{r}_i) $, $T^{\alpha \beta} = Q^{\alpha \beta} + \frac{1}{2} \rho \delta^{\alpha \beta} + \frac{1}{v_0}\sum_i n_i^{\alpha} F_i^{\beta} \delta(\mathbf{r}-\mathbf{r}_i)$, and  $\mu^\alpha$ is a white noise with correlation $\langle \mu^{\alpha}(\mathbf{r}, t) \mu^{\beta}(\mathbf{r}', t')\rangle = \delta^{\alpha \beta} \delta(t - t') \delta(\mathbf{r} - \mathbf{r}')$. From now on, we will focus on the deterministic dynamics of the fields. The effect of noise will be studied elsewhere. 


\subsection{Phenomenological closure}\label{sec: approx}

\begin{figure}
    \centering
    \includegraphics[width=1.0\columnwidth]{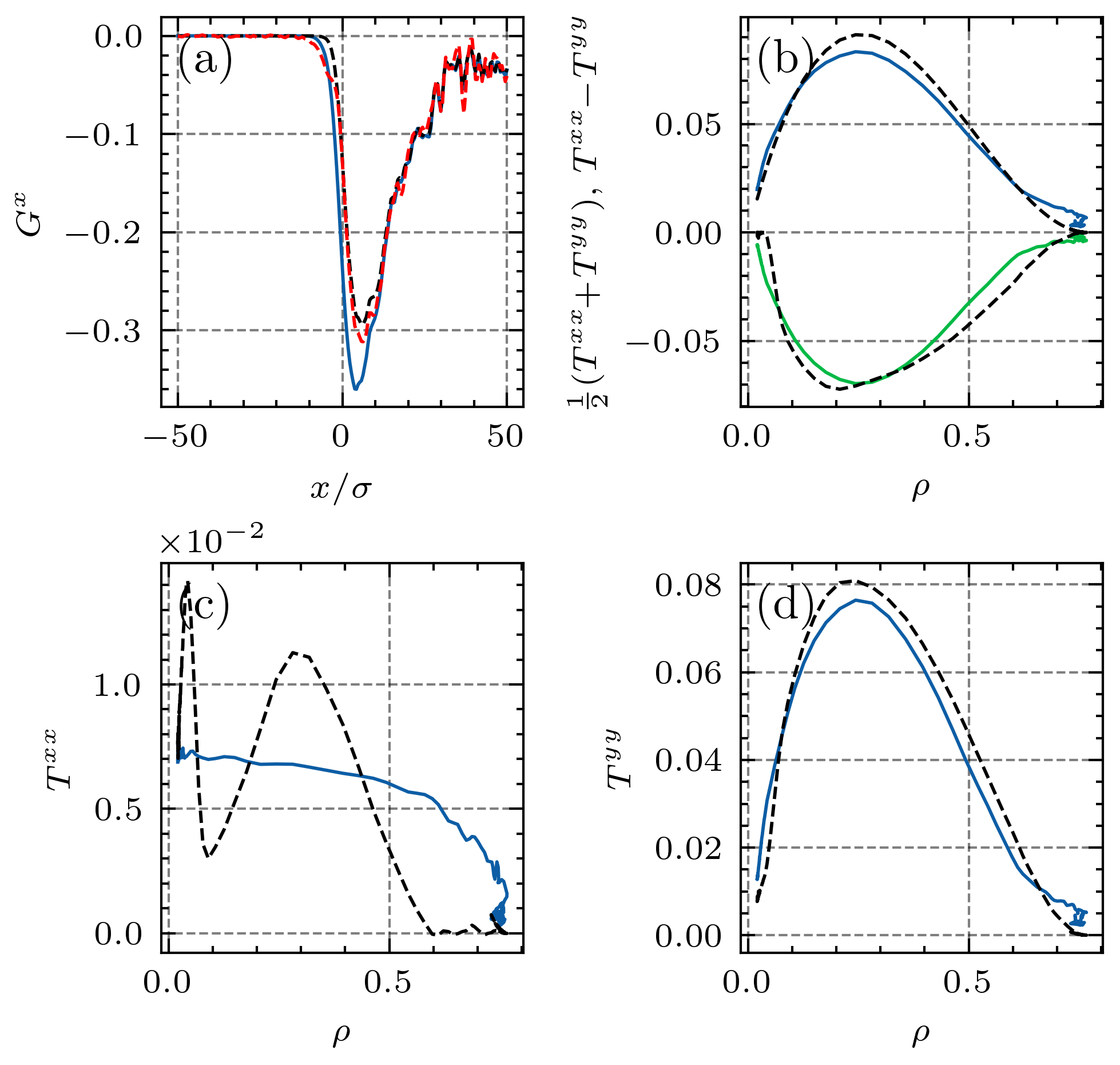}
    \caption{Panel (a) shows the stationary profile of $G^{x}$  with the fit given by Eq.~\eqref{eq: G_approx}. The black (red) dashed line is for the approximation with (without) the thermodynamic term. For the free energy density it is used $f(\rho) = (\rho - \rho_l / 2)^{4}$. Panel (b) shows $(T^{xx} + T^{yy})/2$ in blue and $T^{xx} - T^{yy}$ in green as a function of density. 
    Panels (c) and (d) show in blue the separate components $T^{xx}$ and $T^{yy}$ as a function of density.
    The dashed black lines show the fits given by Eq.~\eqref{eq: T_approx}.}
\label{fig:Tensor T} 
\end{figure}

To obtain a closed set of equations for the fields, it is necessary to approximate $G^{\alpha}$ and $T^{\alpha \beta}$ in terms of the two fields. 
For $G^{\alpha}$, we note that Eq.~\eqref{eq:rho_micro} is a conservation equation with particle flux $J^{\alpha} = v_0(q^{\alpha} + G^{\alpha})$. Since $G^{\alpha}$ depends on interparticle forces, this term is interpreted as the contribution to the current arising from particle interactions. Phenomenologically, this contribution gives rise to two main effects. First, it produces an effective drag that opposes particle motion. This drag increases with density and eventually cancels the self-propulsion at the liquid density $\rho_l$. Such effect can be approximated by a term of the form $-\frac{\rho}{\rho_l} q^{\alpha}$. At $\rho = \rho_l$, this contribution exactly cancels the current due to self-propulsion, $v_0 q^{\alpha}$. This is related to the effective velocity reduction proposed in MIPS theory~\cite{cates_motility-induced_2015}.
Second, in high density zones, particles tend to form a homogeneous state~\cite{hermann_phase_2019}. This can be understood with the simple case where there is a vacancy in zones where $\rho \approx \rho_l$. Due to excluded volume interactions, particles rearrange themselves so that the vacancy disappears. We approximate this effect with a thermodynamic term of the form $-\nabla_{\alpha} \frac{\delta \mathcal{F}}{\delta \rho}$, where $\mathcal{F}$ is a coarse-grained free energy.
Now $G^{\alpha}$ reads,
\begin{equation}
    G^{\alpha} = -\frac{\rho}{\rho_l} q^{\alpha} - \nabla_{\alpha} \frac{\delta \mathcal{F}}{\delta \rho}.
\label{eq: G_approx}
\end{equation}
This approximation is shown in Fig.~\ref{fig:Tensor T}(a) in stationary state. It can be seen that it fits well. 

For the approximation of $T^{\alpha \beta}$, we first note that by symmetry, the off-diagonal components $T^{xy}$ and $T^{yx}$ vanish in this geometry, as the simulations indeed confirm (not shown). The diagonal components are analyzed summing and subtracting them
\begin{align}
    T^{xx} + T^{yy} &= \rho + \frac{1}{v_0}\sum_i \left[q^{x}_i F^{x}_i + q^{y}_i F^{y}_i \right], \label{eq: sum_T}\\
    T^{xx} - T^{yy} &= Q^{xx} - Q^{yy} + \frac{1}{v_0} \sum_i \left[q^{x}_i F^{x}_i - q^{y}_i F^{y}_i \right] \label{eq: res_T}.
\end{align}
Now, similar to the approximation of $G^{\alpha}$, the interparticle forces oppose the particle movement. We write $F^{\alpha}_i \approx -F(\rho_i) n^{\alpha}_i$, where $F(\rho_i)$ is the force magnitude evaluated at the position of particle $i$ and, $F(\rho)$ is a function that increases with density. In this way, $\sum_i q^{\alpha}_i F^{\alpha}_i\approx -\sum_i \left[F(\rho_i) n_i^{\alpha} n_i^{\alpha} \delta(\mathbf{r} - \mathbf{r}_i)\right] \approx -F(\rho) \left[ Q^{\alpha \alpha} + \rho / 2 \right]$. Now Eqs.~\eqref{eq: sum_T} and~\eqref{eq: res_T} read,
\begin{align}
    T^{xx} + T^{yy} &\approx \rho\left( 1 - \frac{F(\rho)}{v_0}\right), \label{eq: sum_T_approx}\\
    T^{xx} - T^{yy} &\approx \left(Q^{xx} - Q^{yy}\right) \left( 1 - \frac{F(\rho)}{v_0}\right)\label{eq: res_T_approx},
\end{align}
where in Eq.~\eqref{eq: sum_T_approx} we used that $Q^{xx}+Q^{yy}=0$.

Eq.~\eqref{eq: sum_T_approx} implies that the trace of $T^{\alpha \beta}$ depends only on the density field. On the other hand, Eq.~\eqref{eq: res_T_approx} implies that the tensor is not isotropic. At the closure level we are working, the nematic tensor is unknown and the anisotropic character of $T^{\alpha \beta}$ should be built in terms of $\mathbf{q}$. In fact, this is approximately true in the polarized regions, where $Q^{xx}\approx q^x q^x/\rho$. With these considerations, we propose the closure 
%
\begin{equation}
    T^{\alpha \beta} \approx¥ h(\rho)\delta^{\alpha \beta} + g(\rho) \left[\frac{q^{2}}{2}\delta^{\alpha \beta} - q^{\alpha}q^{\beta}\right],
    \label{eq: T_approx}
\end{equation}
which give that $T^{xx} + T^{yy} = 2 h(\rho)$ and $T^{xx} - T^{yy} = g(\rho) (q^{x}q^{x} - q^{y}q^{y})$, similar to the conditions given by Eqs.~\eqref{eq: sum_T_approx} and~\eqref{eq: res_T_approx}. Note that we are implicitly describing the nematic tensor in terms of $q^{\alpha}$, so this description cannot capture the tangential contributions that are observed only at nematic order [see Fig.~\ref{fig:fields}(h)]. These contributions could be approximated using density gradients terms of the form $\nabla_{\alpha} \rho q^{\beta}$, $\nabla_{\alpha} \rho \nabla_{\beta} \rho$, or $\nabla_{\alpha \beta} \rho$.

Using our simulation data, averaged in the steady state ($t v_0 / \sigma > 2500$), we measure $T^{\alpha \beta}$ and find the best functions $h(\rho)$ and $g(\rho)$.  From now on, we focus on the case where the fields depend only on the $x$-axis, as in the simulations, and we refer to  $q^x$ simply as $q$. The components of $T^{\alpha \beta}$ with their fits are presented in Fig.~\ref{fig:Tensor T}, where the profiles are aligned to the interface as described in Sec.~\ref{sec.simulations}. We found  $h(\rho) = a \rho(1 - \rho / \rho_0)^{2}$ and $g(\rho) = b [1 - (\rho / \rho_0)^2] / \rho^{3}$, with  $a = 0.4$, $b = 0.1$ and $\rho_0 = \rho_l = 0.77$. 

Equations \eqref{eq:rho_micro} and \eqref{eq:q_micro} with the approximations for $G^{\alpha}$ and $T^{\alpha \beta}$ finally read
\begin{align}
    \partial_{t} \rho &= -v_0 \nabla_x [(1 - \rho / \rho_l) q] 
    + v_0 \nabla_{x}^{2}\frac{\delta \mathcal{F}}{\delta \rho},
    \label{eq:rho_approx} \\
    \partial_{t} q &= -D_r q 
    -v_0 \nabla_x \Big[ {a} \rho (1 - \rho / \rho_0)^{2} \nonumber \\
    &\qquad\quad - b(1 - [\rho / \rho_0]^2) \frac{q^{2}}{2 \rho^{3}} \Big]
    - v_0 \frac{\delta \mathcal{F}}{\delta q},
    \label{eq:qx_approx}
\end{align}
with the effective free energy
\begin{multline}
    \mathcal{F}[\rho, q] = \int dx \bigg[ f(\rho) + \frac{\gamma}{2} (\nabla_x \rho)^{2} +
    \frac{\lambda}{2} (\nabla_x^{3} q)^{2} \bigg].
    \label{eq:eff_energy}
\end{multline}

Here, $f(\rho)$ is a bulk free energy density and the terms with $\gamma$ and $\lambda$ were added to avoid numerical instabilities. Here $\gamma$ penalizes density gradients. For the polarization, the numerical solutions show that a squared gradient is not sufficient to suppress small wavelength instabilities, and a higher order term is added, with coefficient $\lambda$. While the free energy density $f(\rho)$ is general and it does impose a phase transition a priori, the currents in Eqs.~\eqref{eq:rho_approx} and \eqref{eq:qx_approx} vanish in high-density zones, giving rise to the phase transition as discussed for MIPS in~\cite{cates_when_2013}.
For $b = 0$ and $\lambda=0$, Eq.~\eqref{eq:qx_approx} is very similar to Eqs.~(18)  in~\cite{bialke_microscopic_2013} and (9) in~\cite{speck_critical_2022}, where the term accompanied by $a$ acts as a density-dependent effective velocity. The new term, proportional to $b$, which is not present in the previous references,  allows us to describe the dynamics of the polarization field in the homogeneous regime. 

In the infinite persistent regime, $D_r \to 0$, the stationary solution of Eq.~\eqref{eq:q_micro} gives $T^{xx} = 0$. According to the model \eqref{eq: T_approx}, the stationary polarization is given by
\begin{equation}
    q^{x} = \pm \sqrt{2h(\rho) / g(\rho)},
    \label{eq: q_stat_persis}
\end{equation}
which is a function of $\rho$ only. The sign depends on the particle orientation. In this way, the model corrects the divergence that appears in the infinite persistent regime. Furthermore, at low densities the fitted expressions for $g$ and $h$ give $q^x\approx\pm2.8\rho^2$. This quadratic dependence implies that the residual polarization in the gas has a collisional origin.

\subsection{Adiabatic slaving of $q$}
A simple solution for $q$ is achieved by slaving the polarization to the density. Neglecting the temporal derivative and the higher order gradient term on Eq.~\eqref{eq:qx_approx}, leads to
\begin{equation}
    q + L \nabla_{x} \left[ h(\rho) - \frac{1}{2}g(\rho) q^{2} \right] = 0, \label{eq:adiabatic}
\end{equation}
where $L = v_0 / D_r$ is the persistence length, which should be compared to the scale of the gradients $W$. Here, to be concrete, we consider the interface region, where $\rho$ has a step-like profile. For $L\gg W$, it results in Eq.~\eqref{eq: q_stat_persis}. For $L \ll W$, a perturbative solution to first-order leads to $q \approx -L\partial_x h(\rho)$, which qualitatively matches other proposals~\cite{speck_critical_2022, bialke_microscopic_2013} and corresponds to a non-zero small polarization only at the interface.  Solutions to Eq.~\eqref{eq:adiabatic}, valid in the entire space, are obtained by matching the limiting solutions.

\subsection{Stress tensor}
We finally note that the density equation \eqref{eq:rho_approx} can be rewritten mechanically as $\partial_t\rho = -\nabla_\alpha\nabla_\beta \sigma_{\alpha\beta}$, which defines the stress tensor with units absorbing the mobility coefficient~\cite{solon_generalized_2018}. Notably, this stress tensor does not coincide with the thermodynamic one that can be derived from the free energy by making virtual deformations on a stationary interface,
\begin{equation}
    \sigma_\text{p}^{\alpha \beta} = {F}\delta_{\alpha \beta} - \nabla_{\alpha}\rho  \frac{\partial {F}}{\partial (\nabla_{\beta}\rho)} + 
 \nabla_{\alpha} q^{\gamma} \frac{\partial {F}}{\partial (\nabla_{\beta} q^{\gamma})}, 
 \label{eq:pas_stress_tensor}
\end{equation}
where $F$ is the expression between braces on Eq.~\eqref{eq:eff_energy} and the p subscript stands for the ``passive'' part. 
A non-variational stress tensor computed as  \begin{equation}
\sigma_\text{a}^{\alpha \beta } = v_0\int^{\mathbf{r}} d \, r'^{\alpha} [1 - \rho(\mathbf{r}') / \rho_l] \, q^{\beta}(\mathbf{r}')
\end{equation}
must be added to the thermodynamic one to obtain Eq.~\eqref{eq:rho_approx}. This active contribution is key to obtaining the phase separation. Hence, the permanent polarization at the interphase generates an excess stress that keeps stable the wetting film.

\section{Numerical solutions of the two-field equations} 
To map the simulations and the continuum equations, note that the microscopic parameters $v_0$ and $D_r$ enter explicitly in Eqs.~\eqref{eq:rho_approx} and~\eqref{eq:qx_approx}. Thus, the same values as in the simulations are used here: $v_0 = 1$ and $D_r = 10^{-3}v_0 / \sigma$. These parameters fix the units of length and time, thereby matching those used in the simulations. For $a$, $b$, and $\rho_0$ the same estimated values are used ($a = 0.4$, $b = -0.1$ and $\rho_0 = \rho_l$).  Eqs.~\eqref{eq:rho_approx} and \eqref{eq:qx_approx} are solved using the pseudo-spectral solver Dedalus~\cite{burns_dedalus_2020}. As initial conditions, we use a homogeneous density with an excess at the center of the domain (Fig.~\ref{fig:fields}d), imitating what happens in the simulations near the wall. The total mass is fixed to match the total number of particles of the simulation. The parameters used are $L_x = 400$, $\rho_l = 0.78$, $\gamma = 1$, $\lambda=1$. Other values of $\gamma$ and $\lambda$ give similar results, except for small values, where numerical instabilities develop.

For the free energy density, we first simply used $f=0$. In this case, a phase separation develops, but numerical instabilities appear when the local density approaches $\rho_l$ (see Appendix~\ref{app.simfail}). To avoid the instabilities several options are possible (see Appendix~\ref{app.simfail}) and we opted for $f(\rho) = \alpha (\rho - \rho_l/2)^{4}$ with $\alpha = 1$. This choice inhibits the system to reach high and vanishing densities, while at the same time it is rather neutral in imposing preferential densities or a phase separation.

The solutions obtained are shown in Figs.~\ref{fig:fields}(d), (e), (f), and (g), in good agreement with the ABP simulations. The transition between the two polarization regimes is recovered and matches the reorientational time $\tau_q$. Also, the equations capture the polarization of the mass transported to the walls, shown as light-colored regions outside the cluster in Fig.~\ref{fig:fields}(e).  

\begin{figure}[htb]
    \centering
    \includegraphics[width=\columnwidth]{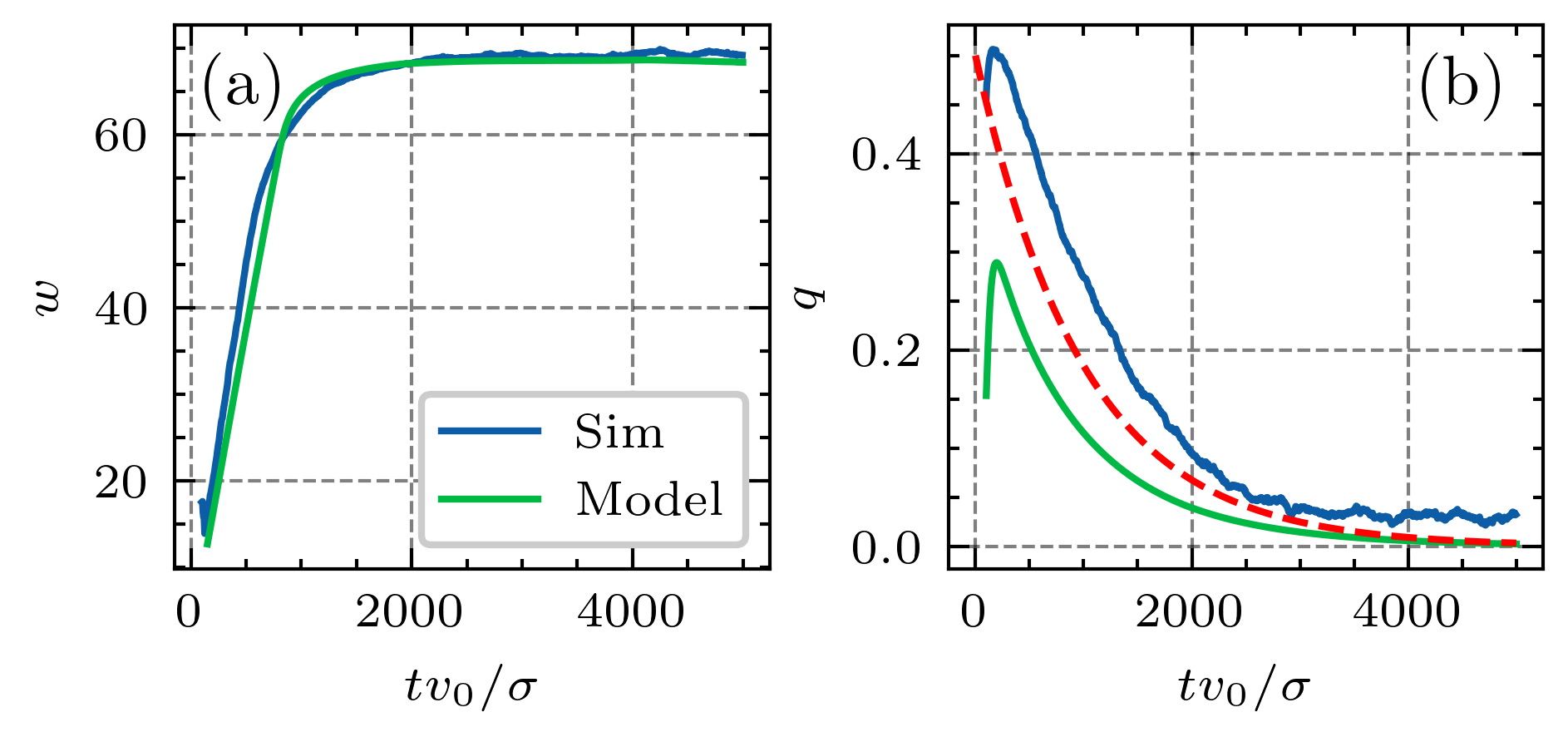}
    \caption{(a) Wetting layer width evolution for simulations and model. (b) Polarization decay near the wall. The dotted line represents the function $\sim\exp(-t / \tau_q)$. }
    \label{fig: width_decay}
\end{figure}

To compare in more detail the solutions obtained by the two field model and the simulations, we study growth of the wetting layer width on time, which is obtained by fitting an hyperbolic tangent to the density profiles. Also, the temporal decay of the polarization in the region near the wall is obtained. These results are shown in Fig.~\ref{fig: width_decay}. 

It can be seen that the growth velocity, which is the slope between times $t = 0$ and $t = 1000$, is nearly identical. Also, the transition between the growth regime and the stationary regime is similar. 

For the polarization decay, it can be seen that there is a difference in the polarization magnitude at short times. This difference could arise from the fact that we approximate \(T^{\alpha \beta}\) with data form the stationary regime. As shown in Eq.~\eqref{eq: q_stat_persis}, the parameters \(a\) and \(b\) are related to the polarization magnitude, so this quantity does not necessarily match the transient regime of the simulations. Nevertheless, for both cases the polarization has a decay with a characteristic time $\sim \tau_q$.

\section{Discussion}
Eqs.~\eqref{eq:rho_approx} and~\eqref{eq:qx_approx} for the density and polarization fields, which were derived by a combination of microscopic and phenomenological arguments, correctly describe the dynamics of a collection of highly persistent ABPs when they phase separate. They are able to account for the transient polarization that emerges during accumulation and also the permanent polarization that establishes at the interphase. The proposed equations can be applied under different configurations when polarization plays a significant role and cannot be simply slaved to density. In particular, they correctly predict finite results even in cases of high persistence.

The model provides a mechanism for the formation of the dense phase purely in terms of the coupled dynamics of density and polarization, without needing to include an effective double well potential in the free energy. Crucially, the mean field approximation used in the MIPS theory, that the effective velocity is reduced when increasing density, also enters in the polarization fluxes $T^{\alpha\beta}$.

Still, there is room for improvement and verification, and we expect that the results in this article will stimulate the research on the dense phases of persistent active matter. First, the expression for $T^{\alpha \beta}$ was obtained from the simulation results at the steady state, and additional terms can appear at the transient. Also, subdominant contributions, proportional to $D_r$ or of higher order in gradients, can play significant roles in specific configurations. For the free energy density $f$, although several functions give qualitatively correct results, more research is needed to fully characterize it~\cite{santos_accurate_1995, marconi_towards_2015}. Finally, the derivation of the equations was done for arbitrary spatial dimensions, but the phenomenological fit for $T^{\alpha \beta}$ was done using simulations that depended only on one dimension. New simulations with more complex spatiotemporal dependencies, as for example those in Refs.~\cite{mangeat_stationary_2024,fins_carreira_how_2024} should be used to fully validate the approximation made or add new terms.

\acknowledgments
This research was supported by the Fondecyt Grant No.~1220536 and Millennium Science Initiative Program NCN19\_170 of ANID, Chile. The authors thank Ignacio Bordeu for discussions on stochastic calculus.

\appendix

\section{Dean method}\label{app.dean}

 Here we present the derivation in details of Eqs.~\eqref{eq:rho_micro} and \eqref{eq:q_micro}.  Starting from the microscopic equations \eqref{eq:vor_abp}, the global and particle density, polarization, and nematic fields are defined as:
\begin{align}
    \rho(\mathbf{r}, t) &= \sum_{i} \rho_i(\mathbf{r}, t) = \sum_i \delta(\mathbf{r}_i - \mathbf{r}),\\
    q^{\alpha}(\mathbf{r}, t) &= \sum_{i} q^{\alpha}_i(\mathbf{r}, t)= \sum_{i}n^{\alpha}_i  \rho_i(\mathbf{r}, t),\\
    {Q}^{\alpha \beta}(\mathbf{r}, t) &= \sum_i Q^{\alpha \beta}_i(\mathbf{r}, t)=\sum_{i} \rho_i(\mathbf{r}, t)\left[ {n}^{\alpha}_i {n}^{\beta}_i  - \frac{\delta^{\alpha \beta}}{2}\right],
\end{align}
where $\alpha$ and $\beta$ indicate the Cartesian components of the vectors and tensors, and the sums run over the particles $i$.
For any differentiable function $f(\mathbf{r})$, it is direct that
$    f(\mathbf{r}_i) = \int \, d\mathbf{r} \rho_i(\mathbf{r}, t) f(\mathbf{r})
$.
A total time derivative of this identity gives
\begin{equation}
\int d\mathbf{r} \, f(\mathbf{r}) \frac{\partial \rho_i(\mathbf{r}, t)}{\partial t}= - \int d \mathbf{r} \, f(\mathbf{r}) \nabla_{\alpha} \left[  \rho_i(\mathbf{r}, t) \frac{d r^{\alpha}_i}{dt}\right].
\end{equation}
Since $f(\mathbf{r})$ is arbitrary, the individual particle densities satisfy:
    $
    {\partial_t \rho_i(\mathbf{r}, t)} = - \nabla_{\alpha} \left[  v_0 q^{\alpha}_i(\mathbf{r}, t) + \rho_i(\mathbf{r}, t) F^{\alpha}_i \right].
    $
Summing over all the particles gives Eq.~\eqref{eq:rho_micro} for the density field 
where $G^{\alpha} = \frac{1}{v_0}\sum_i \rho_i F^{\alpha}_i=\frac{1}{v_0}\sum_i \delta(\mathbf{r}-\mathbf{r}_i) F^{\alpha}_i$. 

For the polarization field, the procedure is similar, but now using
$
    n_i^{\alpha} f(\mathbf{r}_i) = \int d\mathbf{r} \, q^{\alpha}_i(\mathbf{r}, t) f(\mathbf{r})
$, which leads to 
\begin{multline}
        \int d\mathbf{r} \, \rho_i(\mathbf{r}, t) f(\mathbf{r}) \frac{d n_i^{\alpha}}{dt}
        + \int d\mathbf{r} \, \rho_i(\mathbf{r}, t) n_i^{\alpha} \frac{d r_i^{\beta}}{dt} \nabla_{\beta} f(\mathbf{r})\\
        =  \int d\mathbf{r} \, \rho_i(\mathbf{r}, t) f(\mathbf{r}) \left[{\partial_t n_i^{\alpha}}  + D_{r} {\partial_{\theta_i}^2 n_i^{\alpha}}\right] \\
        -  \int d\mathbf{r} \, f(\mathbf{r}) \nabla_{\beta} \left[\rho_i(\mathbf{r}, t) n_i^{\alpha} \frac{d r_i^{\beta}}{dt} \right]. 
\end{multline}
For the first integral, we used Ito's lemma: $d n_i^{\alpha} = \partial_t n_i^{\alpha} dt +  D_r \partial^{2}_{\theta_i} n_i^{\alpha} dt$~\cite{gardiner_stochastic_2009}. 

From the director definition, it follows that ${\partial_t n^{\alpha}_i} = t^{\alpha}_i \sqrt{2 D_r}\eta_i(t)$ and ${\partial_{\theta_i}^{2} n_i^{\alpha}} = - n_i^{\alpha}$, where  $\mathbf{t}_i = ( -\sin \theta_i, \cos \theta_i)$ is the unit vector perpendicular to $\mathbf{n}_i$. By the same arguments used for the density field, the resulting equation is 
$\partial_t q_i^{\alpha} = - D_r q_i^{\alpha} -  \nabla_{\beta} \left( v_0 Q_i^{\alpha \beta} + \frac{v_0}{2} \rho_i \delta^{\alpha \beta} + q_i^{\alpha} F_i^{\beta} \right) + \sqrt{2 D_r} \eta_i t_i^{\alpha}\rho_i$. Summing over all particles gives for the global polarization field 
\begin{equation}
    {\partial_t q^{\alpha}} = - D_r q^{\alpha} -  v_0\nabla_{\beta} T^{\alpha \beta} + \xi^{\alpha},
    \label{eq:q_micro_sm}
\end{equation}
with $T^{\alpha \beta} = Q^{\alpha \beta} + \frac{1}{2} \rho \delta^{\alpha \beta} + \frac{1}{v_0}\sum_i q_i^{\alpha} F_i^{\beta}$. Here,  $ \xi^{\alpha}(\mathbf{r}, t) = \sqrt{2 D_r} \sum_i \eta_i t_i^{\alpha}\rho_i$, which, being the sum of white noises, is also white. Its correlation function is
$
    \langle \xi^{\alpha}(\mathbf{r}, t), \xi^{\beta}(\mathbf{r}', t') \rangle = 2D_r \delta(t - t') \delta^{\alpha \beta} \sum_i \rho_i(\mathbf{r}, t) \rho_i(\mathbf{r}', t) = 2 D_r \rho(\mathbf{r}, t)\delta^{\alpha \beta} \delta(t - t') \delta(\mathbf{r}- \mathbf{r}'),
$ 
where we used that 
$
    \sum_i \rho_i(\mathbf{r}, t) \rho_i(\mathbf{r}', t')  = \rho(\mathbf{r}, t) \delta(\mathbf{r} - \mathbf{r}'),
$
This allows us to define the white noise $\mu^{\alpha}(\mathbf{r}, t)$ with correlation $\langle \mu^{\alpha}(\mathbf{r}, t) \mu^{\beta}(\mathbf{r}', t')\rangle =  \delta^{\alpha \beta} \delta(t - t') \delta(\mathbf{r} - \mathbf{r}')$, such that ${\xi}^{\alpha}(\mathbf{r}, t)=\sqrt{2 D_r\rho(\mathbf{r}, t)} {\mu}^{\alpha}(\mathbf{r}, t)$, which finally gives Eq.~\eqref{eq:q_micro} for the polarization field.

\section{Numerical solution with different $f(\rho)$}\label{app.simfail}

\begin{figure}[htb]
    \includegraphics[width=\columnwidth]{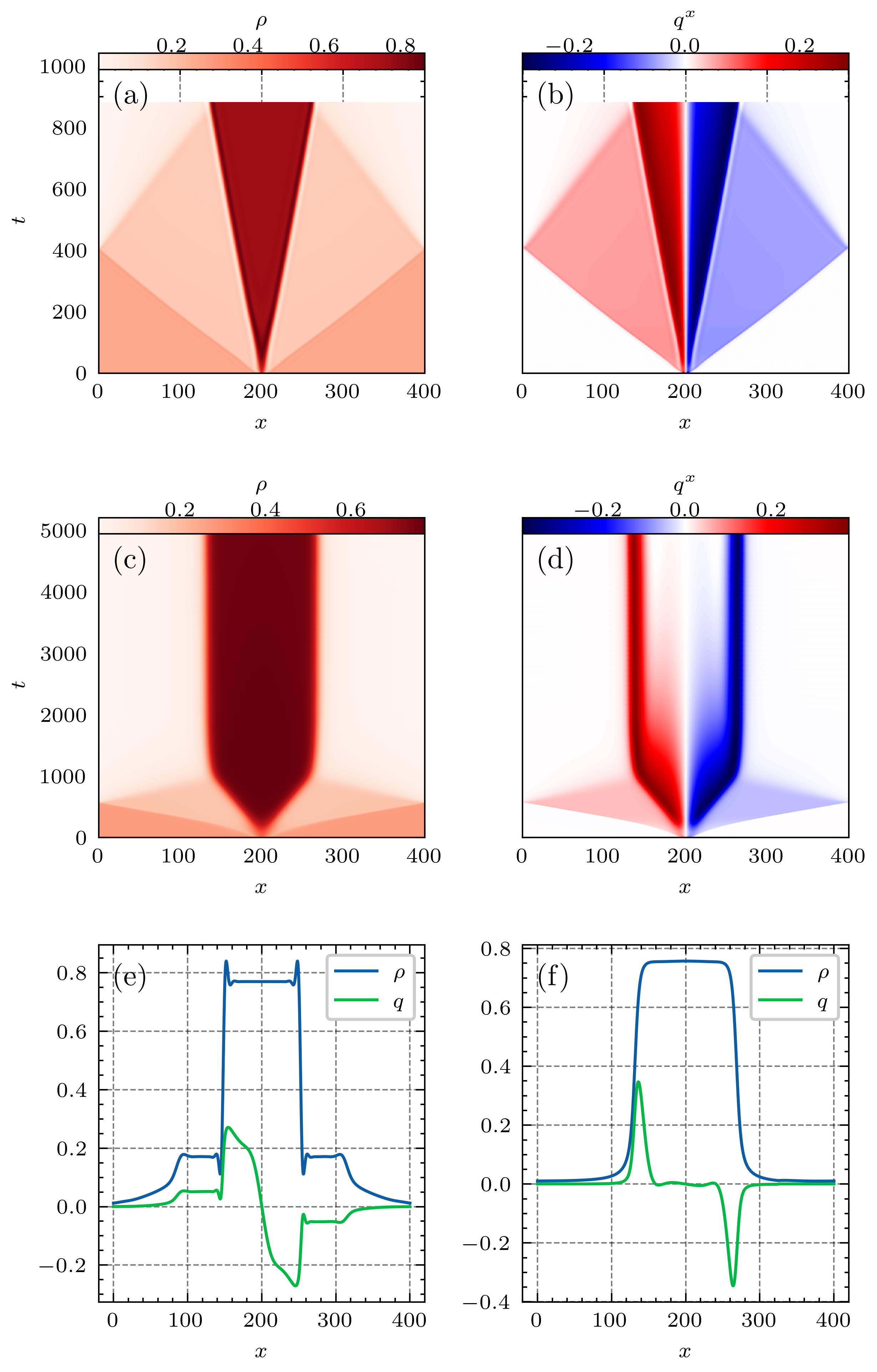}
    \caption{Spatiotemporal diagrams for $\rho$ and $q^{x}$ obtained by solving Eqs.~\eqref{eq:rho_approx} and~\eqref{eq:qx_approx} with different free energy densities. Top: $f(\rho) = 0$. Center: $f(\rho) = (\rho - \rho_l/2)^{2}$.
    Density and polarization profiles (e) for $f(\rho)=0$ (before the instability) and (f) for $f(\rho) = (\rho - \rho_l/2)^{2}$ (stationary state).}
    \label{fig:eqs_sols_f}
\end{figure}

The numerical solutions of Eqs.~\eqref{eq:rho_approx} and~\eqref{eq:qx_approx} with $f(\rho) = 0$ are shown in Figs.~\ref{fig:eqs_sols_f}(a) and (b). The rest of the parameters are the same as in the main text. It can be seen that the phase separation starts to take place, but a numerical instability develops at a finite time, which is seen as an incomplete spatio-temporal diagram. Just before the instability, the density presents a maximum near the interphase that exceeds the liquid density $\rho_l$, as shown in Fig.~\ref{fig:eqs_sols_f}(e). We hypothesize that this may the origin of the numerical instability. Also, this excess density at the interface is not observed in the ABPs simulations.
Free energy densities with barriers at low or high densities, $f(\rho)= \alpha/\rho^n$ or $f(\rho)= \alpha /(\rho_l - \rho)^n$, also result in the initial development of the phase separation to later end due to numerical instabilities. 
A simple quadratic free energy density of the form $f(\rho) = \alpha (\rho - \rho^{*})^{2}$ succeeds in reproducing the phase separation without numerical instabilities. The solutions of 
Eqs.~\eqref{eq:rho_approx} and~\eqref{eq:qx_approx} with $\alpha = 1$ and $\rho^{*} =  \rho_l/2$ are shown in Figs.~\ref{fig:eqs_sols_f}(c) and (d). These results are similar to the ABPs simulations, but with a greater interfacial width, because of the large thermodynamic force at finite densities, which tends to excessively homogenize the system to the central density $\rho^*$. 
Finally, the free energy density $f(\rho) = \alpha (\rho - \rho_l/2)^{4}$, used in the main text, exerts a small thermodynamic force at moderate densities and correctly reproduces the sharper interfaces obtained in the ABPs simulations.

%

\end{document}